\documentclass[aps,twocolumn]{revtex4}%
\usepackage{amssymb}
\usepackage{amsmath}
\usepackage{graphicx}
\usepackage{amsfonts}%
\begin{document}
\preprint{Draft C 9-4-2003}
\title{An RF-Driven Josephson Bifurcation Amplifier for Quantum Measurements}
\author{I. Siddiqi, R. Vijay, F. Pierre, C.M. Wilson, M. Metcalfe, C. Rigetti, L.
Frunzio, and M.H. Devoret}
\affiliation{Departments of Applied Physics and Physics, Yale University, New Haven,
Connecticut 06520-8284}

\begin{abstract}
We have constructed a new type of amplifier whose primary purpose
is the readout of superconducting quantum bits. It is based on the
transition of an RF-driven Josephson junction between two distinct
oscillation states near a dynamical bifurcation point. The main
advantages of this new amplifier are speed, high-sensitivity, low
back-action, and the absence of on-chip dissipation. Pulsed
microwave reflection measurements on nanofabricated Al junctions
show that actual devices attain the performance predicted by
theory.

\end{abstract}

\startpage{1}
\endpage{2}
\maketitle

Quantum measurements of solid-state systems, such as the readout
of superconducting quantum bits
\cite{RFSET,Cottet,Boulder,DELFT,Buisson,LukensSquid,Zorin},
challenge conventional low-noise amplification techniques.
Ideally, the amplifier for a quantum measurement should minimally
perturb the measured system while maintaining sufficient
sensitivity to overcome the noise of subsequent elements in the
amplification chain. Additionally, the characteristic drift of
materials properties in solid-state systems necessitates a fast
acquisition rate to permit measurements in rapid succession. To
meet these inherently conflicting requirements, we propose to
harness the sensitivity of a dynamical system - a single RF-driven
Josephson tunnel junction - tuned near a bifurcation point. The
superconducting tunnel junction is the only electronic dipolar
circuit element whose non-linearity remains unchanged at arbitrary
low temperatures. As the key component of present superconducting
amplifiers \cite{SIS, PARAMP, SQUID}, it is known to exhibit a
high degree of stability. Moreover, all available degrees of
freedom in the dynamical system participate in information
transfer and none contribute to unnecessary dissipation resulting
in excess noise. The operation of our Josephson bifurcation
amplifier is represented schematically in Fig. 1. The central
element is a Josephson junction whose critical current $I_{0}$ is
modulated by the input signal using an application-specific
coupling scheme (input port), such as a SQUID loop \cite{DELFT} or
a SSET \cite{Cottet}. The junction is driven with an sinusoidal
signal $i_{RF}\sin(\omega t)$ fed from a transmission line through
a directional coupler (drive port). In the underdamped regime,
when the drive frequency $\omega$ is detuned form the natural
oscillation frequency $\omega_{p}$ and when the drive current
$I_{\bar{B}}<i_{RF}<I_{B}\ll I_{0}$, the system has two possible
oscillation states which differ in amplitude and phase
\cite{Dykman,Siddiqi}. Starting in the lower amplitude state, at
the bifurcation point $i_{RF}=I_{B}$ the system becomes infinitely
sensitive, in absence of thermal and quantum fluctuations, to
variations in $I_{0}$. The energy stored in the oscillation can
always be made larger than thermal fluctuations by increasing the
scale of $I_{0}$, thus preserving sensitivity at finite
temperature. The reflected component of the drive signal, measured
through another transmission line connected to the coupler (output
port), is a convenient signature of the junction oscillation state
which carries with it information about the input signal. This
arrangement minimizes the back-action of the amplifier since the
only fluctuations felt at its input port arise from the load
impedance of the follower amplifier, which is physically separated
from the junction via a transmission line of arbitrary length and
can therefore be thermalized efficiently to base temperature. In
this Letter, we present an experiment that demonstrates the
principle of bifurcation amplification, and we show that the
sensitivity obtained is in good agreement with a well-established
theory of dynamical transitions, thus making it relevant for
stringent tests of superconducting qubits and gates.

\begin{figure}[b]
\includegraphics[width=3.1in]{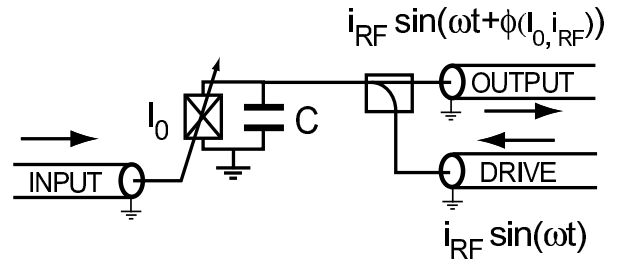}\caption{Schematic diagram of the
Josephson bifurcation amplifier. A junction with critical current $I_{0}$
coupled to the input port is driven by an RF\ signal. In the vicinity of the
dynamical bifurcation point, the phase of the resulting oscillation depends
critically on the input signal and manifests itself in the reflected signal
phase $\phi$.}%
\label{FigSampleWL}%
\end{figure}\qquad

The dynamics of the junction are descibed by the time evolution of
the junction gauge invariant phase difference $\delta\left(
t\right) =\int_{-\infty}^{t}dt^{\prime}2eV(t^{\prime})/\hbar$
where $V$ is the voltage across the junction. In the presence of a
microwave drive $i_{RF}\sin(\omega t)$, the oscillations in the
junction phase can be parametrized using in-phase and quadrature
phase components $\delta(t)=\delta_{\parallel}\sin(\omega
t)+\delta_{\perp}\cos(\omega t)$ (higher harmonics of oscillation
are negligible). When the detuning $\alpha=(1-\omega/\omega_{p})$
and the quality factor $Q=\omega_{p}RC$ satisfy $\alpha
Q>\sqrt{3}/2$, then two steady state solutions can exist for
$\delta(t)$. Here $\omega_{p}=\left( 2eI_{0}/\hbar C\right)
^{1/2}$ is the junction plasma frequency, $C$ is the capacitance
shunting the Josephson element and $R=50\,\Omega$ is the
characteristic impedance of the transmission line at the output
port. The two oscillation states appear as two points in the
$\left( \delta_{\parallel},\delta_{\perp }\right)  $ plane and are
denoted by vectors labelled 0 and 1 (see Fig. 2). The
error-current \cite{Kautz} which describes the generalized force
felt by the system is also plotted in $\left(
\delta_{\parallel},\delta_{\perp }\right)  $ plane. Its value goes
to zero at the attractors corresponding to states 0 and 1 and also
at a third extremum which is the dynamical saddle point. Also
shown in Fig. 2 are the calculated escape trajectory
\cite{Dykmanescape} from state 0 (dashed) and the corresponding
retrappping trajectory \cite{Dmitriev} into state 1 (solid line).
Fig. 2 has been calculated for $\alpha=0.122$, $Q=20$ and
$i_{RF}/I_{B}=0.87$, where $I_{B}=16/(3\sqrt{3})$
$\alpha^{3/2}(1-\alpha)^{3/2}\,I_{0}$ $+\,O(1/(\alpha Q)^{2})$.
These values correspond to typical operating conditions in our
experiment. The dynamical switching from state 0 to 1 is
characterized by a phase shift given here by
$\mathrm{tan}^{-1}\left[  \left(  \delta_{\parallel
}^{1}-\delta_{\parallel}^{0})/(\delta_{\perp}^{1}-\delta_{\perp}^{0}\right)
\right]  =-139\,\mathrm{deg}$. Using the junction phase-voltage
relationship and the transmission line equations, we can calculate
the steady-state magnitude and phase of the reflected microwave
drive signal. The change in the oscillation of $\delta$ results in
a shift of the reflected signal phase
$\Delta\phi_{01}=89\,\mathrm{deg}$. Since there is no source of
dissipation in the junction chip, there should be no change in the
magnitude of the reflected signal power, even though
$\sqrt{(\delta_{\parallel}^{1}-\delta_{\parallel
}^{0})^{2}+(\delta_{\perp}^{1}-\delta_{\perp}^{0})^{2}}\neq0$.

\begin{figure}[b]
\includegraphics[width=3.1in]{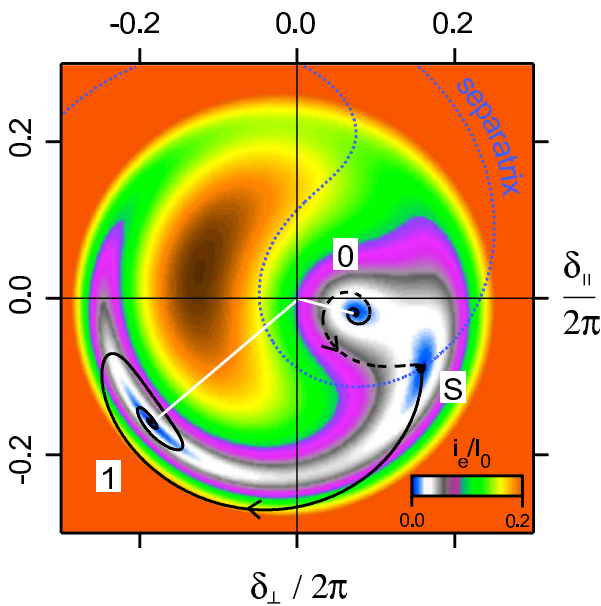}\caption{Poincar\'{e} section of an
RF-driven Josephson junction in the bistable regime
$(\alpha=(1-\omega/\omega_{p})=0.122,i_{RF}/I_{B}=0.87)$. The
coordinates $\delta_{\parallel}$ and $\delta_{\perp}$ are the
in-phase and quadrature-phase components of the junction
gauge-invariant phase difference $\delta$. The color code gives
the magnitude of the error current $i_{e}$ which describes the
"force" on $\delta$. The two stable oscillation states, labeled by
0 and 1, are indicated by white line segments. The basins of
attraction corresponding to the two states are separated by the
blue dotted line (separatrix). Point S which lies on the
separatrix is the saddle point at which the escape trajectory from
state 0 (dashed line) meets the retrapping
trajectory into state 1 (solid line). }%
\label{FigSampleWL}%
\end{figure}\qquad

Our main sample consisted of a single shadow-mask evaporated Al/Al$_{2}$%
0$_{3}$/Al tunnel junction with $I_{0}=1.17\,\mathrm{\mu A}$,
shunted with an on-chip lithographic capacitance
$C=27.3\,\mathrm{pF}$ \cite{Siddiqi} to obtain a reduced plasma
frequency $\omega_{p}/2\pi=1.80\,\mathrm{GHz}$. The dynamics of
the transition between the two oscillation states were probed
using microwave pulses, generated by the amplitude modulation of a
CW source with a phase-locked arbitrary waveform generator with
$1\,\mathrm{ns}$ resolution. The reflected signal was passed
through an isolator at base temperature $T=0.28\,\mathrm{K}$ to a
matched HEMT amplifier at $T=4.2\,\mathrm{K}$ in the spirit of Day
\textit{et al.} \cite{Zmuidzinas}. At room temperature, the
reflected signal was further amplified, mixed down to
$100\,\mathrm{MHz}$ and finally digitally demodulated using a
$2\,\mathrm{GS/s}$ digitizer to extract the signal phase $\phi$.

\begin{figure}[t]
\includegraphics[width=3.1in]{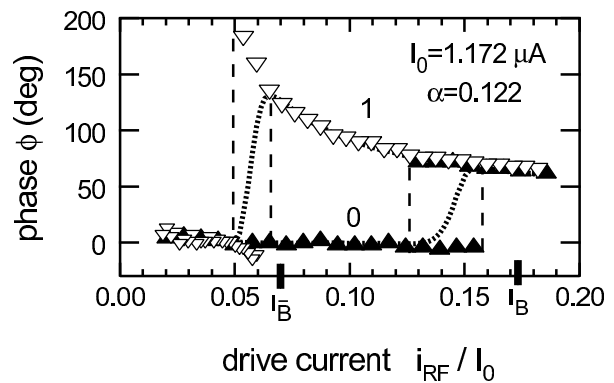}\caption{Reflected signal phase $\phi$
as a function of drive current $i_{RF}/I_{0}$. Symbols denote the mode of
$\phi$ with up and down triangles corresponding to increasing and decreasing
current, respectively. The dotted line is $\left\langle \phi\right\rangle $.
The calculated bifurcation points, $I_{\bar{B}}$ and $I_{B}$, are marked on
the horizontal axis. The 0 and 1 phase states are reminiscent of the
superconducting and dissipative states of the DC current biased junction.}%
\label{FigSampleWL}%
\end{figure}

We first probed the drive current dependence of the reflected
signal phase $\phi\left(  i_{RF}\right)  $ by applying a
$4\,\mathrm{\mu s}$ long symmetric triangular shaped pulse with a
peak value $0.185\,I_{0}$. The demodulated reflected signal was
divided into $20\,\mathrm{ns}$ sections, each yielding one
measurement of $\phi$ for a corresponding value of $i_{RF}$. The
measurement was repeated $6\times10^{5}$ times to obtain a
distribution of $\phi(i_{RF})$. In Fig. 3, the mode of the
distribution is plotted as a function of $i_{RF}/I_{0}$. For
$i_{RF}/I_{0}<0.125$, the bifurcation amplifier is always in state
0, $\phi$ is constant and assigned a value of
$0\,\mathrm{deg}$. As the drive current is increased above $i_{RF}%
/I_{0}=0.125$, thermal fluctuations are sufficiently large to
cause transitions to the 1 state. In the region between the two
dashed lines at $i_{RF}/I_{0}=0.125$ and $i_{RF}/I_{0}=0.160$,
$\phi$ displays a bimodal distribution with peaks centered at $0$
and $74\,\mathrm{deg}$ with the latter corresponding to the
amplifier in the 1 state, as we have demonstrated previously
\cite{Siddiqi}. The dotted line in Fig. 3 is the average reflected
signal phase $\left\langle \phi\right\rangle $. When
$i_{RF}/I_{0}$ is increased above $0.160$, the system is only
found in state 1. In the decreasing part of the $i_{RF}$ ramp, the
system does not start to switch back to state 0 until
$i_{RF}/I_{0}=0.065$. The critical switching currents $I_{B}$ for
the $0\rightarrow1$ transition and $I_{\bar{B}}$ for the
$1\rightarrow0$ transition, calculated from numerical simulations
to treat the inductance of wire bonds, are denoted with ticks in
Fig. 3, and are in good agreement with experiment.  The hysteresis
$I_{\bar{B}}<I_{B}$ is a consequence of the asymmetry in the
escape barrier height for the two states. Thus, the
$0\rightarrow1$ transition at $i_{RF}=I_{B}$ is nearly
irreversible, allowing the bifurcation amplifier to latch and
store its output during the integration time set by the
sensitivity of the follower amplifier.

\begin{figure}[b]
\includegraphics[width=3.1in]{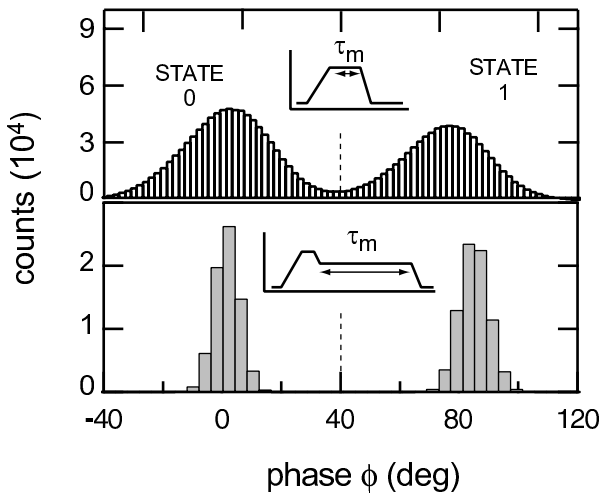}\caption{Histograms of the reflected
signal phase $\phi$ at $i_{RF}/I_{0}=0.145$. The upper histogram
contains $1.6\times10^{6}$ counts with a measurement time
$\tau_{m}=20\,\mathrm{ns}$. The lower panel, taken with the
latching technique, has $1.5\times10^{5}$ counts with a
measurement time $\tau_{m}=300\,\mathrm{ns}$. Data here has been
taken under the same operating conditions as in Fig 3. The dashed
line represents the discrimination threshold between the 0 and 1 state.}%
\label{FigSampleWL}%
\end{figure}

To determine the sensitivity of the bifurcation amplifier, we have
characterized in detail the switching in the vicinity of the
$0\rightarrow1$ transition. We excited the system with two
different readout pulse protocols. In the first protocol, the
drive current was ramped from 0 to its maximum value in
$40\,\mathrm{ns}$ and was then held constant for $40\,\mathrm{ns}$
before returning to 0. Only the final $20\,\mathrm{ns}$ of the
constant drive period were used to determine the oscillation phase
with the first $20\,\mathrm{ns}$ allotted for settling of the
phase. Histograms taken with a $10\,\mathrm{MHz}$ acquisition rate
are shown in Fig. 4. In the upper panel, the two peaks
corresponding to states 0 and 1 can easily be resolved with a
small relative overlap of $10^{-2}$. The width of each peak is
consistent with the noise temperature of our HEMT amplifier. In
this first method, the latching property of the system has not
been exploited. In our second protocol for the readout pulse, we
again ramp for $40\,\mathrm{ns}$ and allow a settling time of
$20\,\mathrm{ns}$, but we then reduce the drive current by $20\%$
and measure the reflected signal for $300\,\mathrm{ns}$. In that
latter period, whatever state was reached at the end of the
initial $60\,\mathrm{ns}$ period is "latched" and time is spent
just increasing the signal/noise ratio of the reflected phase
measurement. As shown in the lower panel of Fig. 4, the two peaks
are now fully separated, with a relative overlap of
$6\times10^{-5}$ allowing a determination of the state 1
probability with an accuracy better than $10^{-3}$. This second
protocol would be preferred only for very precise time-resolved
measurements of $I_{0}$ or for applications where a low-noise
follower amplifier is impractical.

\begin{figure}[t]
\includegraphics[width=3.1in]{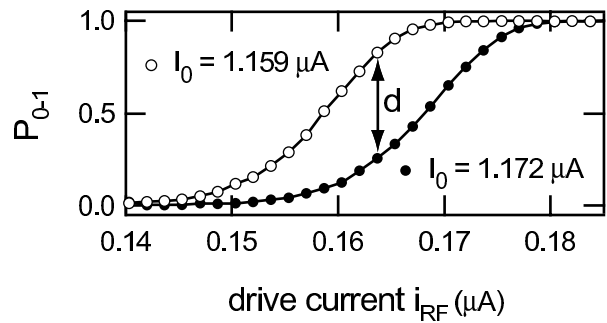}\caption{Switching probability curves
at $T=280\,\mathrm{mK}$ as a function of the drive current $i_{RF}$. The
discrimination power $d=0.57$ is the maximum difference between the two
curves. The measurement protocol is the same as shown in the upper panel of
Fig. 4. Though the two curves differ by approximately $1\%$ in $I_{0}$, they
are shifted by $6\%$ in drive current.}%
\label{FigSampleWL}%
\end{figure}

A third experiment was performed to study the state 1 switching
probability $P_{0\rightarrow1}\left(  i_{RF}\right)  $ for
different values of the temperature $T$ and $I_{0}$, the latter
being varied with a magnetic field applied parallel to the
junction plane. Using the first readout protocol and the
discrimination threshold shown in Fig. 4, we obtain the switching
probablity curves shown in Fig. 5. Defining the discrimination
power $d$ as the maximum difference between two switching
probability curves which differ in $I_{0}$ we find that at $T=280\,\mathrm{mK}%
$, $d=57\%$ for $\Delta I_{0}/I_{0}=1\%$ -- the typical variation observed in
a superconducting charge-phase qubit \cite{Vion}. The switching probability
curves should shift according to $(\Delta I_{B}/I_{B})/(\Delta I_{0}%
/I_{0})=3/(4\alpha)-1/2+O(1/(\alpha Q)^{2})$, which for our case
takes the value 5.6. In Fig. 5, the curves are shifted by $6\%$,
which agrees well with this prediction. For the case of the DC
current biased junction, similar curves would shift only by $1\%$
since the switching current is $I_{0}$ itself. Comparable
discrimination power using DC switching has only been achieved in
these devices at $T\lesssim60\,\mathrm{mK.}$ As the temperature is
increased, the switching probabilility curves broaden due to
increased thermal fluctations and the disciminating power
decreases: at $T=480\,\mathrm{mK}$, $d=49\%$.

We finally determined the escape rate $\Gamma_{0\rightarrow1}(i_{RF}%
,I_{0},T)=\omega_{a}/2\pi\,\exp\left(  -\Delta U_{dyn}^{0}/k_{B}T\right)  $ as
a function of $i_{RF}$ by measuring the time dependence of the switching
probability, using a method previously applied to the determination of the
static switching rates to the voltage state \cite{Turlot}. Here $\omega_{a}$
is the attempt frequency and the barrier height can be approximated as $\Delta
U_{dyn}^{0}=\,u_{dyn}^{0}(1-(i_{RF}/I_{B})^{2})^{3/2}$ where $u_{dyn}%
^{0}=64\hbar/(18e\sqrt{3})\,\,I_{0}\,\,\alpha(1-\alpha)^{3}$. After the
initial ramp ($40\,\mathrm{ns}$) and settling period ($20\,\mathrm{ns}$), the
reflected signal phase was extracted every $20\,\mathrm{ns}$ for a duration of
$1\mathrm{\mu s}$. By repeating this measurement, we generated switching
probability histograms which we analyzed as $P_{0\rightarrow1}\left(
t\right)  =1-\exp(-\Gamma_{0\rightarrow1}t)$. $\ $ In parallel with these
dynamical switching measurements, we have run static switching measurements
with the same analysis, yielding the escape temperature $T_{st}^{esc}$. We
have found that $T_{st}^{esc}$ exceeded $T$ by 60mK, an effect we do not fully
understand yet but which might be attributed to a combination of insufficient
filtering in our RF amplifier line outside the measurement band and to
high-frequency non-equilibrium vortex motion in the superconducting films of
our device. Analyzing the dynamical switching data with $T_{st}^{esc}$ in
place of $T$, we extract a value of $u_{dyn}^{0}=10.7\,\mathrm{K}$ from the
$T=280\mathrm{mK}$ data with $I_{0}=1.17\,\mathrm{\mu A}$ while the calculated
value keeping higher order terms in $1/\alpha Q$ is $u_{dyn}^{0,calc}%
=11.0\,\mathrm{K}$.

With the Josephson bifurcation amplifier operating at $T_{st}^{esc}%
=340\,\mathrm{mK}$, it is possible to resolve with a signal/noise ratio of 1 a
$1\%$ variation in $I_{0}$ in a total time $\lesssim80$ $\,\mathrm{ns}$,
corresponding to a critical current sensitivity of $S_{I_{0}}^{1/2}%
/I_{0}=2.8\times10^{-6}\,\mathrm{Hz}^{-1/2}$. This value is in
agreement with the approximate prediction
$S_{I_{0}}^{1/2}\propto[\partial(\Delta U^{0}_{dyn}/kT)/\partial
I_{0}]^{-1}\cdot\tau_{m}^{1/2}$. This formula shows that the
sensitivity is independent of $Q$, in contrast with the
sensitivity obtained in the linear regime \cite{Delft2}. Operating
the junction near a bifurcation point has the advantage that the
sensitivity is limited only by temperature and not by the
measurement bandwidth. The advantage of the bifurcation amplifier
over SQUIDs \cite{Clarke} resides in its extremely low
back-action. Since there is no on-chip dissipation, the only
source of back-action is the junction phase fluctuations induced
by our matched isolator load at $T=280\,\mathrm{mK}$.$\ $Both
numerical simulations and analytical
calculations predict a spectral density $S_{\delta}\approx\,10^{-12}%
\,\mathrm{rad}^{2}/\,\mathrm{Hz}$. In our system, the generalized
back-action force is $f(t)=-\hbar/2e\,\cos\delta(t)$ and we
calculate a noise temperature
$T_{N}=\sqrt{S_{f}S_{I_{0}}}/k_{B}\lesssim10\,\mathrm{mK}$,
considerably lower than that of resistively shunted SQUIDs where
$T_{N}$ $\approx T$ at comparable temperatures \cite{Urbina}.
Finally the bifurcation amplifier does not suffer from
quasiparticle generation associated with hysteretic SQUIDS
\cite{DELFT} and DC\ current-biased junctions \cite{Cottet} which
switch into the voltage state. Long quasiparticle recombination
times at low temperatures limit the acquisition rate of these
devices while the recombination process itself produces excess
noise for adjacent circuitry \cite{Lukens}.

In conclusion, the Josephson bifurcation amplifier is competitive
with the SQUID for applications where low noise temperature is
required. Its speed, suppression of on-chip dissipation, and
latching make it ideal for the readout of superconducting qubits.
At temperatures such that $T_{dyn}^{esc}\leq60\,\mathrm{mK}$, the
discrimination power would be greater than 95\%, hence permitting
stringent tests of Quantum Mechanics, like the violation of Bell's
inequalities.

We would like to thank D. Prober, E. Boaknin, L. Grober, D.
Esteve, D. Vion, S. Girvin and R. Schoelkopf for discussions and
assistance. This work was supported by the ARDA (ARO Grant
DAAD19-02-1-0044) and the NSF (Grant DMR-0072022).

\end{document}